\documentclass[11pt,twoside]{article}

\usepackage{asp2006}
\usepackage{epsf}
\usepackage{psfig}

\begin{document}

\title{Winds of Main-Sequence Stars:
Observational Limits and a Path to Theoretical Prediction}
\author{Steven R. Cranmer}
\affil{Harvard-Smithsonian Center for Astrophysics,
60 Garden Street, Cambridge, MA 02138, USA}

\begin{abstract}
It is notoriously difficult to measure the winds of solar-type
stars.  Traditional spectroscopic and radio continuum techniques
are sensitive to mass loss rates at least two to three orders of
magnitude stronger than the Sun's relatively feeble wind.  Much
has been done with these methods to probe the more massive outflows
of younger (T Tauri) and older (giant, AGB, supergiant) cool stars,
but the main sequence remains terra incognita.  This presentation
reviews the limits on traditional diagnostics and outlines more
recent ideas such as Lyman alpha astrospheres and charge-exchange
X-ray emission.  In addition, there are hybrid constraints on mass
loss rates that combine existing observables and theoretical models.
The Sackmann/Boothroyd conjecture of a more massive young Sun (and
thus a much stronger ZAMS wind) is one such idea that needs to be
tested further.  Another set of ideas involves a strong proposed
coupling between coronal heating and stellar mass loss rates, where
the former is easier to measure in stars down to solar-like values.
The combined modeling of stellar coronae and stellar winds is
developing rapidly, and it seems to be approaching a level of
development where the only remaining ``free parameters'' involve the
sub-photospheric convection.  This talk will also summarize these
theoretical efforts to predict the properties of solar-type
main-sequence winds.
\end{abstract}

\section{Introduction}

All stars are believed to possess expanding outer atmospheres known
as stellar winds.
Continual mass loss has a significant impact on the evolution of
the stars themselves, on surrounding planetary systems, and on the
overall mass and energy budget of the interstellar medium (see
reviews by Dupree 1986; Lamers \& Cassinelli 1999; Willson 2000).
We would like to understand how stellar wind properties depend on
the fundamental stellar parameters and to identify the physical
processes that drive the winds.
Observations are key, of course, and recent advances in measuring
the plasma properties of our own {\em solar wind} (via both
remote sensing and in-situ probes) have been helpful.
There is a well-known mismatch, though, between the most basic
property of the solar wind (it's mass loss rate:
$\dot{M} \approx 10^{-14} \, M_{\odot} \, \mbox{yr}^{-1}$) and
corresponding values for detectable outflows from other stars
($\dot{M} > 10^{-10} \, M_{\odot} \, \mbox{yr}^{-1}$,
typically).
Solar-type winds are just too tenuous to make much of
an impact on star-integrated photon signatures.

The goal of this paper is to summarize a cross-section of
observational and theoretical work that is improving our
understanding of the low-density winds from solar-type
(cool, main-sequence) stars.

\begin{figure}[!ht]
\plotone{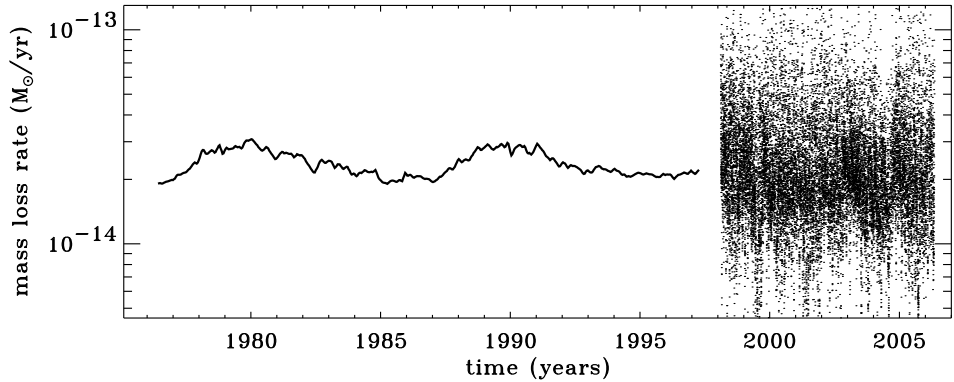}
\caption{Time variation of solar mass loss rate computed in
two ways: Wang's (1998) full-Sun reconstruction ({\em solid
curve, 1976--1997}), and instantaneous mass fluxes from the
product of in-situ densities and velocities, from the SWEPAM
instrument on {\em ACE} ({\em points, 1998--2006}).}
\end{figure}

\section{The Solar Wind}

The last decade has seen significant progress toward identifying
and characterizing the processes that heat the solar corona and
accelerate the solar wind (see, e.g., Cranmer 2002, 2004;
Kohl et al.\  2006; Aschwanden 2006; Klimchuk 2006).
It seems increasingly clear that {\em closed} magnetic loops in
the low corona are heated by small-scale, intermittent magnetic
reconnection that is driven by the continual stressing of their
footpoints by convective motions.
The {\em open} field lines that reach into interplanetary space,
though, appear to be energized by the dissipation of waves and
turbulent motions (also driven ultimately by convection).
Parker's (1958) classical paradigm of solar wind acceleration via
the gas pressure gradient in a hot ($T \sim 10^{6}$ K) corona still
seems to be the dominant mechanism, though waves and turbulence
have an impact as well.
Clues about the detailed process of turbulent heating have come
from UV spectroscopy of the extended corona (i.e., using a
combination of an occulting coronagraph and a spectrometer).
There is evidence for preferential acceleration of heavy ions in
the fast solar wind, ion temperatures exceeding $10^8$ K, and
marked departures from Maxwellian velocity distributions---all
of which point to specific types of {\em collisionless kinetic}
processes at the end of the turbulent cascade.

The Sun-integrated mass loss rate $\dot{M}$ is a quantity that
is not often considered by solar physicists, since both
telescopic and in-situ measurements typically resolve much
smaller volumes of plasma at a time.
It is possible to use in-situ measurements of the density $\rho$
and flow speed $v$ to estimate a time-dependent mass flux
$\dot{M} = 4\pi \rho v r^{2}$ (as if the entire heliosphere
is identical to the measured parcel).
Average values tend to fall around
$2 \times 10^{-14} \, M_{\odot} \, \mbox{yr}^{-1}$, with
the majority of data points falling within about a factor
of 3 of this value (i.e., the $\pm 2 \sigma$ range is between
0.6 and $6 \times 10^{-14} \, M_{\odot} \, \mbox{yr}^{-1}$);
see Figure 1.
Wang (1998) used some well-known empirical correlations between
the plasma properties measured at 1~AU and the coronal magnetic
field geometry (the latter extrapolated from photospheric
fields) to reconstruct a sphere-integrated value of $\dot{M}$
over two solar cycles.
He found this ``true'' mass loss rate to vary only
by about 50\%, from 2 to
$3 \times 10^{-14} \, M_{\odot} \, \mbox{yr}^{-1}$, with
lower [higher] values at solar minimum [maximum].

The Sun's mass flux is generally believed to be determined
in the thin {\em transition region} between the chromosphere and
corona by a balance between downward conduction (from the corona),
upward enthalpy flux (from the chromosphere), and local radiative
losses (e.g., Hammer 1982; Withbroe 1988; Leer et al.\  1998;
Leer \& Marsch 1999).
Essentially, {\em more coronal heating implies more mass loss.}
Alternate ideas, such as Alfv\'{e}n wave pressure acceleration
(Holzer et al.\  1983) and a dynamic balance between ionization
and recombination in the chromosphere (Peter \& Marsch 1997),
have an impact on the mass flux but do not seem to be the
dominant drivers.
From a modeling perspective, it has been realized that it is key
to resolve the full range of radii---from the nearly hydrostatic
chromosphere to the supersonic wind---in order to simulate a
self-consistent flow along the open flux tube.

\section{Cool-Star Mass Loss Rates}

Figure 2 shows a summary of $\dot{M}$ determinations from a wide
variety of stars.
The sources of these data are given below; when $T_{\rm eff}$ or
$L_{\ast}$ were not provided, they were determined from
spectral types and luminosity classes using the relations of
de Jager \& Nieuwenhuijzen (1987).
Uncertainty limits are not shown in Figure 2, but it is clear that
the sources of potential error increase dramatically as the methods
grow increasingly {\em model-dependent.}
By necessity, this occurs as the derived values of $\dot{M}$
decrease.
Note also that Figure 2 omits observations of ``bursty''
mass loss episodes (implied for, e.g., some M-dwarf flare stars;
Mullan et al.\  1992) as well as jets associated with interacting
binaries; the goal is to include only time-steady mass loss
from individual stars.

\begin{figure}[!ht]
\plotone{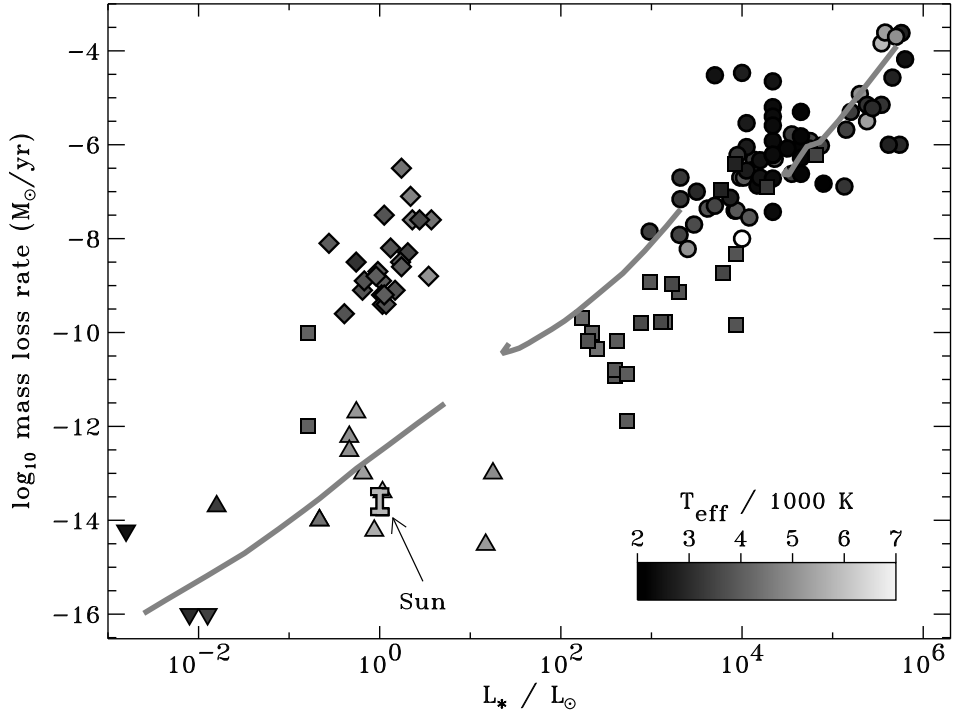}
\caption{Compilation of cool-star mass loss rates, plotted
against stellar luminosity (abscissa) and effective temperature
(symbol grayscale).
{\em Circles:} evolved stars (de Jager et al.\  1988),
{\em diamonds:} T Tauri stars (Hartigan et al.\  1995),
{\em squares:} newer results from spectroscopy (see text),
{\em up-pointing triangles:} nearby stars with astrospheres
(Wood et al.\  2002, 2004, 2005),
{\em down-pointing triangles:} pre-CV M dwarfs (Debes 2006).
The measured range of the Sun's mass loss rate is shown with a
thick error bar, and semi-empirical rates from the
Schr\"{o}der \& Cuntz (2005) law are shown with gray curves
(left to right: luminosity classes V, III, and Iab).}
\end{figure}

The circles in Figure 2 correspond to the de Jager et al.\  (1988)
database of evolved stars, with mass loss rates derived largely
from four main techniques:
\begin{enumerate}
\item
{\em Optical and UV spectroscopy:}
Blueshifted absorption lines probe the column of outflowing gas
in front of the stellar disk.
If the wind is dense enough, there may also be a line-centered
emission component indicative of ``off-limb'' expansion;
i.e., a P Cygni profile.
If the wind is too tenuous, the line opacity is dominated by the
more-or-less static photosphere.
For most lines, deriving $\dot{M}$ is possible only in combination
with a model atmosphere having an assumed ionization balance.
\item
{\em IR continuum excess:}
For cool stars, circumstellar dust gives rise to a measurable
continuum enhancement in the infrared.
The gas-to-dust ratio and the outflow speed must be assumed in
order to obtain $\dot{M}$.
\item
{\em Molecular lines:}
For the densest outflows
($\dot{M} > 10^{-7} \, M_{\odot} \, \mbox{yr}^{-1}$), molecular
emission lines (from, e.g., CO, OH, CN, CS) at mm and sub-mm
wavelengths have provided accurate outflow speeds and mass loss
rates in the ``shells'' far from the parent stars.
Sometimes enough information is present to piece together
details about the mass loss histories of some stars---i.e.,
more than just the first derivative of $M_{\ast}(t)$.
\item
{\em Radio continuum excess:}
Ionized plasma in an outflowing atmosphere emits in the radio 
via bremsstrahlung.
The ionization state and the outflow speed must be assumed in
order to obtain $\dot{M}$.
\end{enumerate}
For more information on these techniques, see references cited
by de Jager et al.\  (1988) and Judge \& Stencel (1991).

In the 1990s, more self-consistent analyses of visible and UV
spectra began to be performed, thanks to both increasingly
sophisticated radiative transfer models and better data (e.g.,
{\em FUSE,} GHRS/{\em{HST,}} and high spectral
resolution ground-based instruments).
The squares in Figure 2 highlight $\dot{M}$ determinations
from Dupree et al.\  (1990, 2005), Harper et al.\  (1995),
Robinson et al.\  (1998), Carpenter et al.\  (1999), and
Lobel \& Dupree (2000).
Also included are several additional values of $\dot{M}$ from
Judge \& Stencel (1991) determined using the above traditional
techniques, which do not overlap with the de Jager et al.\  (1988)
database (and are not chemically peculiar).
The diamonds in Figure 2 illustrate outflows from T~Tauri
stars as determined from optical forbidden line profiles
(Hartigan et al.\  1995).

More recently, several other new diagnostic techniques have been
explored that attempt to push the lower limit of detectability
down past the solar mass loss rate.
Wood et al.\  (2002, 2004, 2005) used high-resolution {\em HST}
spectra of H~I Ly$\alpha$ lines from nearby stars to characterize
the properties of their so-called ``astrospheres.''
The interaction between a fully ionized stellar wind and the
partially ionized local interstellar medium (LISM) gives rise
to a dense concentration of neutral hydrogen atoms between the
bow shock and astropause.
This ``hydrogen wall'' ($T \approx 30,000$ K) is hotter than the
exterior LISM gas ($T \approx 8000$ K) and thus shows up as an
absorption component on the blue side of the H~I Ly$\alpha$ line.
The density in the hydrogen wall is proportional to the wind's
mass loss rate, and Wood et al.\  (2002, 2004, 2005) have
calibrated this diagnostic using multi-fluid hydrodynamic models.
The up-pointing triangles in Figure 2 show the derived mass loss
rates from Wood et al.\  (2004).
These $\dot{M}$ determinations are are somewhat controversial,
though, because they depend sensitively on:
(1) astrosphere models that cannot be verified for the individual
stars, and (2) modeled interstellar absorption in the core and
wings of the intrinsic (and unobserved) stellar Ly$\alpha$ profile.
Improvements in the dynamical and radiative transfer models should
help reduce the uncertainties.

Another new technique for determining the mass loss rates of
some M dwarfs in pre-cataclysmic-variable binary systems was
suggested by Debes (2006).
Some H-rich white dwarfs show metal lines in their
atmospheric spectra that would be difficult to maintain
(against downward diffusion) if the metal-rich gas were not
being continually accreted.
Debes (2006) computed the accretion rates by comparing the
observed strengths of Ca~II H and K lines to models that predict
the balance between mass addition and downward Ca ``settling.'' 
The inter-binary density is then computed by assuming the
accretion takes place via standard Bondi-Hoyle transonic flow,
and this density is then used to compute the companion star's
mass loss rate.
The wind speed $v$ is an assumed quantity in this analysis,
and the derived mass loss rate depends sensitively on $v^4$.
The down-pointing triangles in Figure 2 show derived $\dot{M}$
values for the three close binaries analyzed by Debes (2006).
The three other (wider) binary systems studied by Debes (2006)
have much larger inferred mass loss rates, but these values
are uncertain to within at least 2 orders of magnitude, and there
are inconsistencies between these values and the evolutionary
histories of these systems that lend doubt to the results.

Finally, there is another suggested technique that at present
has yielded only upper limits on mass loss rates.
Wargelin \& Drake (2001, 2002) suggested that the winds of
nearby dwarf stars may be detectable via X-rays induced by
charge exchange.
Interstellar hydrogen flows into stellar astrospheres and
undergoes charge exchange with ions in the supersonic wind;
these ions are left in an excited state and emit X-rays.
Because the charge exchange rate is proportional to the product
of the ions' density and velocity, the X-ray emission is
sensitive to the stellar mass loss rate.
With good enough spatial and spectral resolution (which may
not be achievable until the {\em Generation-X} mission) this
diagnostic can also be used to constrain the wind's velocity,
elemental composition, and ionization state.

\section{Coronal Heating: A Link Between Theory and Observation?}

Theoretical explanations for cool-star mass loss tend to follow
various components of the {\em energy flux} from the stellar
interior up through the atmosphere.
Reimers (1975, 1977) assumed the stellar wind luminosity (or
wind power) remains proportional to the bolometric photon
luminosity as cool stars evolve; i.e., that
\begin{equation}
  \frac{L_{\rm wind}}{L_{\ast}} \, \approx \,
  \frac{\dot{M} V_{\rm esc}^2}{L_{\ast}} \, \approx \,
  \frac{GM_{\ast} \, \dot{M}}{L_{\ast} R_{\ast}} \, \approx \,
  \mbox{constant.}
\end{equation}
Schr\"{o}der \& Cuntz (2005) obtained a similar relation by
assuming that the wind's energy flux is set by the turbulent
transfer of subphotospheric convective energy to photospheric
wave motions (see the curves in Figure 2 for a comparison
between the observed $\dot{M}$ values and those from the
Schr\"{o}der \& Cuntz scaling law).
Models of wave generation from turbulent convection have been
used to model chromospheric and coronal heating in the Sun
and other stars.

The connections between coronal heating and stellar winds can be
utilized to obtain a better empirical understanding of mass
loss.
The primary observational signature of coronal heating---X-ray
emission---is seen in a much larger fraction of cool stars
than those that have measured mass loss rates.
As models of the solar corona and wind improve in sophistication
and become less dependent on free parameters, it may be possible
to extend them to other stars and determine (albeit still in a
model-dependent way) consistent ranges of wind properties
for stars with measured X-ray fluxes.

Taking the assumption that coronal heating causes mass loss,
there appear to be two primary roadblocks in the way of producing
reliable models for solar-type stars:
\begin{enumerate}
\item
Mass loss proceeds along {\em open} magnetic flux tubes, but
the dominant X-ray emission comes from {\em closed} magnetic
loops.
For very active stars, is there any hope of extracting the X-ray
signature of just the ``dim'' open regions?
If the heating in the open regions is related causally to the
(more observable) heating in the closed regions, though, the
latter may scale with the former.
A combination of diagnostics---not just X-rays---appears to be
needed in order to put separate constraints on the open and closed
magnetic regions.\footnote{%
The situation is even more complicated for T~Tauri stars, for
which the open flux tubes may contain either outflowing
wind or inflowing accretion material (see, e.g.,
Feigelson \& Montmerle 1999; Matt \& Pudritz 2005;
Dupree et al., these proceedings).}
\item
For the open flux tubes, there is still disagreement in the solar
community on the extent that coronal heating is driven by
``magnetic activity'' (i.e., simply proportional to some positive
power of the magnetic flux at the base of the corona) or by
the wave fluxes at the stellar surface (presumably determined
by convection).
Cranmer et al.\  (2007) produced a model of chromospheric and
coronal heating along open flux tubes that reproduces a wide
range of observations of fast and slow solar wind streams;
these models depend on a {\em combination} of the magnetic
and wave parameters (see also
Schwadron \& McComas 2003; Suzuki 2006a).
Rapidly rotating active stars appear to have stronger
magnetic fields than the Sun (e.g., Saar 2001) and they may also
have more vigorous convection (K\"{a}pyl\"{a} et al.\  2006;
Brown et al.\  2007).
\end{enumerate}
An interesting testbed for the above issues is the age dependence
of solar-type stars---i.e., the ``Sun in time.''
Figure 3a shows a selection of X-ray luminosities (in units of
the stars' bolometric luminosities) as a function of age.
The $L_{\rm X}$ values were either measured in the ROSAT/PSPC-like 
0.1--2.4 keV band or converted into that band.
Values from individual stars were taken from
G\"{u}del et al.\  (1998) and Garc\'{\i}a-Alvarez et al.\  (2005);
the cluster means come from Preibisch \& Feigelson (2005) and
Jeffries et al.\  (2006).
The solar range of values (error bar at $t = 4.6$ Gyr) is
from Judge et al.\  (2003).
The fit shown with a dashed line is given by
\begin{equation}
  \frac{L_{\rm X}}{L_{\rm bol}} \, = \, 
  \frac{4.48 \times 10^{-4}}{1 + (12.76 \, t_{\rm Gyr})^{1.79}}
  \,\, .
  \label{eq:Lxfit}
\end{equation}

\begin{figure}[!ht]
\plotone{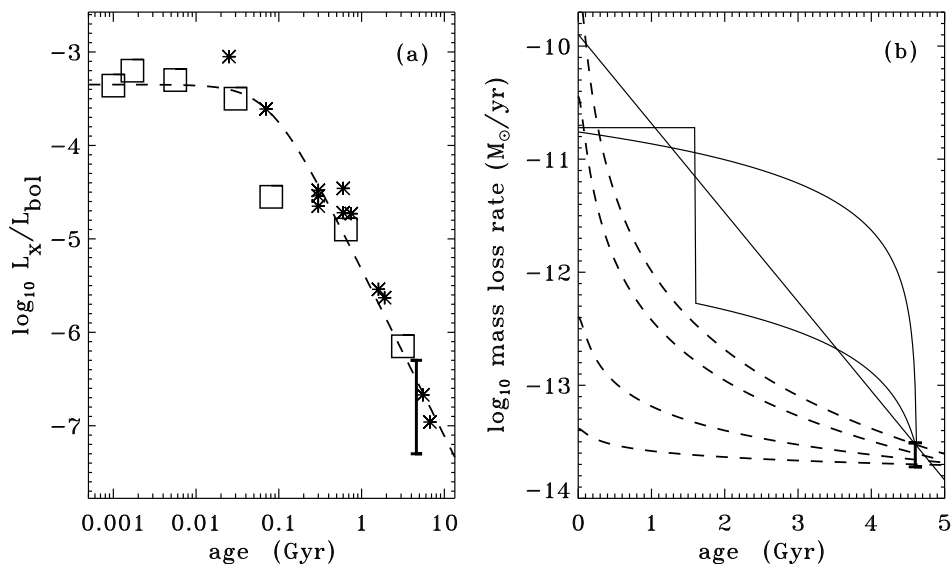}
\caption{(a) X-ray luminosity versus age for solar-type stars
({\em asterisks}) and nearby clusters ({\em squares});
see text for details.
(b) Modeled solar mass loss history, anchored to the
present-day range, where dashed curves assume
$\dot{M} \propto (L_{\rm X}/L_{\rm bol})^{\mu}$, with
$\mu = 0.1,$ 0.4, 1.0, 1.3 (bottom to top), and solid
curves are the ``optimal'' modeled histories from
Sackmann \& Boothroyd (2003).}
\end{figure}

There have been several attempts to scale mass loss rates with
stellar X-ray fluxes $F_{\rm X}$ (or luminosities $L_{\rm X}$)
and with magnetic field fluxes $\Phi$.
The latter two quantities are well-known to be nearly linearly
correlated for the Sun, over the solar cycle, as well as for
a range of other cool stars; i.e.,
$F_{\rm X} \propto \Phi^{\alpha}$, with
$\alpha \approx 0.9$--1.1 (e.g., Pevtsov et al.\  2003;
Schrijver et al.\  2003).
Astrospheric mass loss rates were used by
Wood et al.\  (2002, 2004, 2005) to correlate $F_{\rm X}$
with $\dot{M}$.
Values of the scaling exponent $\dot{M} \propto F_{\rm X}^{\mu}$
ranged from $\mu \approx 1$ to 1.3 depending on the subset of
stars used in the correlation.
The Wood et al.\  database was analyzed in a different way by
Holzwarth \& Jardine (2007), who found a weaker relationship
between $F_{\rm X}$ and $\dot{M}$ ($\mu \approx 0.5$) by
constructing fast-magnetic-rotator type models of the winds
and separating out several K dwarfs from the sample that appear
to have anomalously high mass loss rates.

Small values of $\mu$ can also be derived by returning to the
solar analogy.
Note that the Sun's total mass loss rate varies relatively weakly
over the solar cycle (i.e., by about 50\%, as computed by
Wang 1998), whereas its magnetic flux varies by at least
two orders of magnitude over the cycle.
Schwadron et al.\  (2006) separated the magnetic flux variation
into ``active region'' (AR) and ``quiet Sun'' (QS) components
based on magnetic field models.
For the years between 1985 and 1997, we cross-correlated the
Wang (1998) mass loss data against the Schwadron et al.\  (2006)
magnetic flux curves.
Assuming $\alpha = 1$ (to convert $\Phi$ into $F_{\rm X}$), we
found $\mu \approx 0.1$ for AR, and $\mu \approx 0.4$ for QS.
In a sense, these are ``apples-and-oranges'' comparisons, since
the closed-field AR and QS regions do not connect directly to
the solar wind.
Schwadron et al.\  (2006) also computed the ``coronal hole''
(i.e., open field) component of the magnetic field, but did not
show the considerably weaker $F_{\rm X}$ or $\Phi$ variations
that would be attributed to these regions.
Presumably these would exhibit $\mu$ closer to unity.

Figure 3b shows a reconstruction of the Sun's mass loss history
by taking eq.~(\ref{eq:Lxfit}) and simply assuming that
$\dot{M} \propto (L_{\rm X}/L_{\rm bol})^{\mu}$ for the range of
$\mu$ estimates discussed above.
Also shown are hypothetical mass loss histories of the Sun
that were constructed by Sackmann \& Boothroyd (2003) to 
solve the so-called ``faint young Sun problem'' (e.g.,
Sonett et al.\  1991) by positing a higher initial mass for
the Sun (1.01--1.07 $M_{\odot}$).
It is difficult to reconcile these mass loss histories with
the (substantially lower) values computed from the above range
of $\mu$ exponents.

However, implicit in the $F_{\rm X}^{\mu}$ scaling laws used
above is the assumption that the strength of the young Sun's
subphotospheric convection is essentially independent of
age and activity.
The validity of this assumption remains to be seen.
Standard stellar evolution models do not include any substantial
rotation dependence in their prescriptions for convection,
but numerical simulations like those of Brown et al.\  (2007)
are beginning to show that rapid rotation creates more
intermittent or ``patchy'' convection with sites of substantially
stronger turbulent motion than the present-day solar case.
Open magnetic field lines that feed the stellar wind are
jostled by waves whose amplitudes scale directly with the
peak convective velocity, and the mass loss rate depends on
the deposition of this wave energy in the upper atmosphere
(e.g., Cranmer et al.\  2007).

\section{Conclusions}

Considerable progress has been made over the last several decades
in increasing our understanding of the physics of cool-star winds.
Much of this progress has come about because of cross-fertilization
between the solar and stellar communities, as well as between
theorists and observers.
The diagnostic techniques described in {\S}~3 for obtaining mass
loss rates continue to be developed and extended as the data
accumulate and improve in quality.
Extrasolar planet studies are helping to accelerate this
observational renaissance as well, since they are resulting in
both a dramatic increase in data {\em quantity} and in new
indirect probes of stellar wind acceleration regions.\footnote{%
There is growing evidence for some kind of magnetic interaction
between stars and close-in extrasolar giant planets (with
star-planet distances less than about 0.1 AU).
Some observed chromospheric enhancements appear to be phased
with the planet's orbit (e.g., Shkolnik et al.\  2005) and
seem to signal flare-like energy releases due to the interaction
between the stellar and planetary magnetic fields.
As these observations improve there is hope that both the
field geometries and the stellar wind properties can be probed
far from the stellar surfaces (Saar et al.\  2004;
Preusse et al.\  2006; Cranmer \& Saar, these proceedings).}

Theoretical models of stellar atmospheres and winds are
advancing rapidly.
Artificial features of past models, such as parameterized
``coronal heating functions'' and wave damping lengths, are
now being replaced by more self-consistent and self-regulating
physical processes (e.g.,
Dmitruk et al.\  2002; Gudiksen 2005; Suzuki 2006a, 2006b;
Cranmer \& van Ballegooijen 2005; Cranmer et al.\  2007).
In order for such models to be applied to a range of cool stars,
though, both the subphotospheric convection dynamics and the
magnetic field strength and geometry must be better constrained.
These properties are difficult to observe directly, but the
ongoing cross-pollination between the observational and
theoretical communities continues to provide new insight.

\acknowledgements This work was supported by NASA under grant
{NNG\-04\-GE77G} to the Smithsonian Astrophysical Observatory.
We thank the {\em ACE} SWEPAM instrument team and the {\em ACE}
Science Center for providing the data used in Figure 1
(http://www.srl.caltech.edu/ACE/ASC/).

\end{document}